\documentstyle[aps,multicol,graphicx]{revtex}

\begin{document}

\title{Correlations in the Sine-Gordon Model
with Finite Soliton Density}

\author{D.N. Aristov \cite{PNPI} and A. Luther}
% \email{}
%\affiliation{%
\address{
NORDITA, Blegdamsvej 17, DK-2100, Copenhagen, Denmark
}%

\date{\today}
\maketitle

\begin{abstract}
We study the sine-Gordon (SG) model at finite densities of the
topological charge and small SG interaction constant, related to
 the
one-dimensional Hubbard model near half-filling.  Using the
modified WKB approach, we find that the spectrum of the Gaussian
fluctuations around the classical solution reproduces the
results of the Bethe ansatz studies. The modification of the
collective coordinate method allows us to write down the action,
free from infra-red divergencies. The behaviour of the
density-type correlation functions is non-trivial and we
demonstrate the existence of leading and sub-leading asymptotes.
A consistent definition of the charge-raising operator is
discussed. The superconducting-type correlations are shown to
decrease slowly at small soliton densities, while the spectral
 weight of right
(left) moving fermions is spread over neighboring  ``$4k_F$''
 harmonics.
\end{abstract}


\begin{multicols}{2} \narrowtext

\section{introduction}

In this paper we study the sine-Gordon (SG) model at finite
densities of the topological charge and small SG interaction
constant.  The exactly solvable sine-Gordon model has a large
number of applications in different subfields of condensed
matter physics and statistical mechanics.  The particular case
of the SG model of our interest below arose previosly in various
studies of the one-dimensional Hubbard model close to
half-filling, \cite{MIT} commensurate-incommensurate transition,
\cite{PokroTal} strongly correlated systems in higher dimensions
\cite{square}, collective excitations in a multi-fluxon
Josephson junction \cite{Lebwohl}, 
superconducting  \cite{Rodriguez} or quantum Hall \cite{Hanna}
double-layer systems in parallel magnetic field 
 and phase excitations in ferroelectric liquid crystals
 \cite{Kutnjak}.

Theoretical investigations of the quantum SG model with the
finite density of topological excitations (kinks) were
undertaken  by Haldane \cite{haldane82},
Caux and Tsvelik, \cite{CauxTsve} Papa and Tsvelik.
\cite{PapaTsve} These studies, employing the analysis of the
Bethe anzats equations, established the dispersion of low-lying
excitations and the so-called Luttinger parameter, defining the
asymptotic behavior of the correlation functions.

In the present work, we use an alternative method
to investigate this problem, which was introduced by
Dashen, Hasslacher and Neveu, \cite{Dashen} and is usually
called the modified WKB or semiclasscal approach. In this method,
one first finds the classical solution to the equation of motion
and then analyzes the quantum fluctuations around this solution.
\cite{raja} It is known that this method gives excellent
agreement with the available exact solutions of several models,
and particularly the SG model.

With the use of the modified WKB approach, we
find that the spectrum of Gaussian fluctuations in our model
gives the spectrum coinciding with the one  obtained earlier by
the Bethe anzats method. \cite{haldane82,PapaTsve} At the same
time, applying the standard semiclassical treatment, we
encounter  strong infra-red divergencies in higher
orders of perturbation theory. The difficulties arise also with
the definition of the so-called dual field associated with the
charge-raising operator of the theory. It is known that a
troublesome difficulty with the semiclassical approach is the
existence of  fluctuations bearing  zero energy, which
corresponds in our case to the translational symmetry of the
problem. The treatment of these ``zero modes'' is best done by
the collective coordinate method \cite{ChristLee}, whose
straightforward application, though, does not cure our
problem.
It is shown below however that it is possible to use a
somewhat modified collective coordinate method, to arrive at the
action, free from the infra-red divergencies.
Technically it is done by allowing the fluctuating field to enter
the argument of the classical solution. The stable infra-red
action is achieved at the expense of a more complicated form of
 the
correlations.

Proceeding this way, we obtain the leading and subleading
asymptotic terms in the density-density correlation functions in
agreement with earlier predictions by Haldane. \cite{haldane82}
The definition of the charge-raising operator now becomes free
from ambiguities and we can calculate the leading behavior of
the superconducting-type correlations. Remarkably, the
decay of the latter correlations is slow at small densities
of kinks, an observation which confirms  earlier results.
\cite{haldane82,PapaTsve}

We demonstrate a non-zero quantum average 
of the four-fermion umklapp operator in the Hubbard model
away from half-filling. Particularly, it results in the 
spectral density of the right-moving fermions 
being distributed over the neighboring $4k_F$ harmonics of
the density. 

The remaining part of the paper is organized as follows :
We obtain and analyze the classical solution in Sec.\
\ref{sec:classic}. The fluctuations around it are discussed in
Sec.\ \ref{sec:fluctu}. The difficulties with the
standard application of the semiclassical approach are outlined
in Sec.\ \ref{sec:diffic}. A way of circumventing these
difficulties is discussed in Sec.\  \ref{sec:solu}, we present
the form of the correlation functions here.
Concluding remarks are found in Sec.\  \ref{sec:conclu}.
The essence of the collective coordinate method is given in
the Appendix.

\section{classical solution}
\label{sec:classic}
We consider the Lagrangian density of the form
        \begin{equation}
        {\cal L}=
        \frac12 (\partial_t\phi)^2 -
        \frac12 (\partial_x\phi)^2 -
        \frac{m^2}{\beta^2}(1-\cos\beta\phi)
        + H \frac{\beta}{2\pi}\partial_x\phi
	\label{lagr00}
        \end{equation}
on a line of length $L$.
The SG interaction constant is assumed to be small, $\beta \ll
 1$, and
$m$ is the first breather mass.
The term with the chemical potential $H$ does not enter the
 equation of
motion and defines the boundary conditions for the field $\phi$.
We write the topological charge density $\rho = \frac1{2\pi}
\partial_x(\beta\phi)$
and the average density $\bar\rho = Q/L$.
Using the well-known correspondence between the fermions and
 bosons in
one spatial dimension, \cite{GNT} we associate $\rho$ with the
 fermionic
density. Particularly,   $\cos\beta\phi$ appears
as an umklapp term in the charge sector of the Hubbard model
and $\bar\rho$ measures the deviation from
half-filling, as $2\pi \bar\rho = 4k_F -2\pi$.
The total charge $Q$ is given by
        \begin{equation}
        Q = \int_0^L \rho \,dx =
        \frac1{2\pi} \beta\phi\left.\right|_0^L,
        \end{equation}
which defines the boundary conditions for $\phi$.

In what follows, we employ the quasiclassical method of analysis
 of the
quantum system defined by (\ref{lagr00}). This method
arises most naturally in path integral formulation of the problem
 and
basically corresponds to the steepest descent method. At a first
 step, one
finds the classical configuration of the field, delivering the
 extremum to
the exponentiated action, and then analyzes the spectrum of
 fluctuations
around it. The general criterion for the applicability of the
 method is
the formally large value of the classical action; it was shown,
 however,
that the quasiclassics gives exact results
for a particular case of sine-Gordon model with a zero density of
 the
topological charge. \cite{Dashen}

For later convenience we make a shift $\beta\phi \to
\beta\phi+\pi$ writing  the potential term in the form
        \begin{equation}
        U(\phi) =
        \frac{m^2}{\beta^2}(1+\cos\beta\phi)
        \end{equation}
As a first step we seek a static classical solution $\phi_0$ by
variating ${\cal L}$ in $\delta\phi$. We have
        \begin{equation}
        \partial_x^2\phi_0 - U'(\phi_0) = 0
        \end{equation}
Multiplying it by $\partial_x\phi_0$ and integrating over $x$ we
 obtain
        \[
        (\partial_x\phi_0)^2 = 2 U(\phi_0) + C
        \]
with some constant of integration $C$, which  is integrated
again to give
        \[
        x = \int^{\phi_0}\frac{dy} {\sqrt{2 U(y) +C}}
        \]
Letting $C = \frac{4m^2}{\beta^2}\frac{1-k^2}{k^2}$ with yet
undetermined $k$, we get
        \[
        x = \frac km F(\beta\phi_0/2,k) + const,
        \]
with incomplete elliptic integral $F$. \cite{Ab-St}
Inverting the last equality, we have
        \begin{equation}
         \phi_0 = 2\beta^{-1} {\rm am}(xm/k,k)
         \label{classic}
        \end{equation}
where ${\rm am}(x,k)$ the Jacobi amplitude function \cite{Ab-St}
with the
elliptic index $k$. This index  is determined by the above  total
 variation
of the field $(\beta\phi/2)$ which results in the equation
 \cite{FvdM}
        \begin{equation}
        \frac m{2\bar\rho} = k K(k)
        \label{eq-k}
        \end{equation}
with the complete elliptic integral $K$.
Henceforth we will omit the index $k$ as a second argument of
the elliptic functions and use
the conventions $k_1 = \sqrt{1-k^2}$, $K(k)=K$, $K(k_1)=K'$,
 $E(k)=E$.
From (\ref{eq-k}) we have
        \begin{eqnarray}
        k_1 &\simeq& 4e^{- m/2\bar\rho}, \quad  \bar\rho \ll m,
 \\
               &\simeq& 1 ,  \quad  \bar\rho \gg m.
        \end{eqnarray}

Introducing the soliton mass $M_s = 8m/\beta^2$ and
rescaling the field $H=M_s h$ we write  the energy of this
classic solution, ${\cal E}_0 = -{\cal L}[\phi_0]$ as follows
        \begin{equation}
        \frac{\beta^2}{2m^2}\frac{{\cal E}_0}{L} =
        \frac{2E-k_1^2 K}{k^2K} - \frac{2h}{kK}
        \label{energy-cl}
        \end{equation}
The index $k$, defining the density (number of kinks),
 should be chosen in order to minimize the energy ${\cal E}_0$.
Differentiating (\ref{energy-cl}) by $k$, we find that the energy
 attains
its minimum at
        \begin{equation}
        {E}/{k}=h
        \label{field-k}
        \end{equation}
and this minimum has a form
        \begin{equation}
        {\cal E}_0 = - \frac{2m^2 L}{\beta^2}
        \frac{k_1^2 }{k^2}.
        \label{energy-min}
        \end{equation}
The last expression coincides, up to the sign, with the pressure
 $p=
-\partial{\cal E}_0/\partial L|_Q $. The enthalpy of the system
 is given by
$
        \frac{2m^2L}{\beta^2}
        \frac{2E}{k^2K}
$.

\subsection{Number of Kinks}
At smaller fields, $h<1$, there are no solutions to Eq.
(\ref{field-k}).
The critical value of the field for the appearance of the ``kink
 condensate'',
$H_{cr}=M_s$ was found earlier in \cite{PapaTsve}.
 At fields, slightly exceeding this critical value, we
have for the density $(m/\bar\rho) e^{-m/\bar\rho}\simeq
 (h-1)/4$. Upon
further increase of the field, we have approximately linear
 relation between
the field and the density of kinks :
        \begin{eqnarray}
        \bar\rho &\simeq&
        m \left( \ln\frac{4H_{cr}}{H-H_{cr}} \right)^{-1}
        , \quad H\simeq H_{cr}
        \label{rho-h1} \\
        &\simeq&
        H\beta^2/(4\pi^2)  , \quad H\gg H_{cr}
         \label{rho-hgg}
        \end{eqnarray}
Numerically, the asymptotic expression for large kink densities
(\ref{rho-hgg}) holds with a good accuracy already at $H\geq
 2H_{cr}$.

\subsection{Susceptibility}
Consider now the susceptibility defined as $\chi=
 -L^{-1}\partial^2
{\cal E}_0/ \partial H^2$.
A simple calculation gives
        \begin{eqnarray}
        \chi
 &=& \frac{\beta^2}{16} \frac{E}{k_1^2K^3}
        \\
        &\simeq&
        \frac{\beta^2}{8(h-1)}
         \left( \ln\frac{8}{h-1} \right)^{-2}
        , \quad h\simeq 1
        \label{chi-h1} \\
        &\simeq&
        (\beta/2\pi)^2  , \quad h\gg 1
         \label{chi-hgg}
        \end{eqnarray}
Our expressions for the kink density $\bar\rho$ and
 susceptibility
$\chi$ are identical with those obtained earlier \cite{PapaTsve}
 in the
large-field limit, eqs.(\ref{rho-hgg}), (\ref{chi-hgg}).
At the fields $H\simeq H_{cr}$ our expressions for $\rho$ and
 $\chi$
coincide with their counterparts in \cite{PapaTsve} up to an
 overall factor
$\pi/2$.

Near its minimum at $\rho=\bar\rho$, the energy can be expanded
 as follows
         \begin{equation}
         {\cal E}_0 \simeq  \frac {L}{2\chi} (\rho-\bar\rho)^2.
         \label{ene-rho}
         \end{equation}
or, $1/\chi = \pi v_N$, in the notations by Haldane.
 \cite{haldane82}

Another important parameter $v_J$, according to Haldane, is
given by the coefficient in the total Hamiltonian ${\cal H} =
v_J P^2/(2\pi L\bar\rho^2)$, at small values of the total field
momentum
         $P = \int_0^L dx\, \dot \phi \phi' $.
It can be easily shown, that ${\cal H} = P^2/2{\cal M}_{tot} $
with the mass of the kinks' condensate ${\cal M}_{tot} =
\int_0^L dx\, (\partial_x\phi_0)^2 = LE/K (2m/\beta k)^2 = Q H$.
As a result, one finds $v_J = \pi \bar\rho /H = \pi \beta^2/(16
K E)$. Therefore, one expects \cite{haldane82} the Fermi
velocity $v_F=\sqrt{v_Nv_J} = k_1K/E$ and the Luttinger
parameter ${\cal K} = \sqrt{v_J/v_N}=\pi\beta^2/(16k_1K^2)$,
both expectations verified below.

\section{fluctuations around the classical solution}
\label{sec:fluctu}
\subsection{Lame Equation}

Now we consider the fluctuations around the classical
solution $\phi_0$.  We write
$\phi=\phi_0+\eta$ and expand the Lagrangian into Taylor series
\cite{raja}

        \begin{eqnarray}
        {\cal L} &=& {\cal L}[\phi_0] + {\cal L}_1 \\
        {\cal L}_1 &=&
        \frac12 (\partial_t\eta)^2 -
        \frac12 (\partial_x\eta)^2 +
        \frac12 m^2 (\cos\beta\phi_0) \eta^2
        \label{L-expa}
        \\&&+ higher\, orders\, in\, \eta
         \nonumber
        \end{eqnarray}
The term linear-in-$\eta$ drops out from this expression and
the higher orders in $\eta$ contain  additional powers of
 $\beta$, and
we will neglect them for the moment.
We look
for the solutions $\eta$ in the form $\eta(x,t) = \sum c_i(t)
\eta_i(x)$, obeying cyclic boundary conditions.  Integrating by
parts, we write
        \begin{eqnarray}
        {\cal L}_1 &=&
        \frac12 \eta\left(-
        \partial_t^2 +
        \partial_x^2 +
        m^2 \cos\beta\phi_0 \right) \eta
        \end{eqnarray}

\noindent
Next, we seek the ``normal modes'', satisfying
        \begin{equation}
        (-\partial_x^2 -
        m^2 \cos\beta\phi_0 ) \eta_i(x) = \omega^2_i \eta_i(x)
        \label{Schro}
        \end{equation}
The dynamics of these normal modes is simple,
$c(t) = e^{\pm i\omega t} c$. Introducing the variable $z=xm/k$
 we rewrite
Eq. (\ref{Schro}) in the form \cite{Lebwohl}
        \begin{equation}
        \frac {d^2}{dz^2} \eta =
        (A + 2k^2 {\rm sn}^2 z) \eta
        \label{eq:Schro}
        \end{equation}
with $A = -k^2(\omega^2/m^2+1)$ and ${\rm sn}(x)$ Jacobian
 elliptic
function.  This is the Jacobi form of the Lam\'e equation and the
 solutions
to it are described in the literature . (Ref. \cite{WW2}, chapter
 23.71)
These solutions
are parametrized by an index $\alpha$ and are explicitly written
 as
        \begin{equation}
        \eta(z,\alpha) = \frac{{\rm H}(z-\alpha)} {\Theta(z)}
        \exp(z {\rm Z}(\alpha))
        \label{eta1}
        \end{equation}
with Jacobi functions ${\rm H}(u) = \vartheta_1(\pi u/(2K))$,
$\Theta(u)=\vartheta_4(\pi u/
(2K))$ and ${\rm Z}(u)=d\ln
\Theta(u)/du$; the nome of the
$\vartheta-$functions $\tilde q=\exp(-\pi K'/K)$. \cite{Ab-St}
The structure of (\ref{eta1}) shows that
${\rm H}/\Theta$ is
the modulating Bloch function, and ${\rm Z}$ corresponds to a
wave vector. The energy $\omega$ is given by
        \begin{equation}
        \omega = \pm (m/k)\,
        {\rm dn}(\alpha)
        \label{eq-omega}
        \end{equation}
The second independent solution for $\eta$ with the same energy
(\ref{eq-omega}) is obtained from
(\ref{eta1}) by changing $\alpha\to -\alpha$ which corresponds to
 the
symmetry of eqs. (\ref{eq:Schro}), (\ref{eta1}) with respect
to reflection $z\to -z$.

\subsection{Properties of $\eta(z)$}

One can show that the whole family of linearly independent
 solutions
(\ref{eta1}), which possess the real-valued energies and are
 periodic on a
length of the chain, is exhausted by the values of $\alpha$
belonging to segments $(K-2iK',  K]$ and $[-2iK',0)$ in the
complex plane. We will refer to these segments as to segment I
and II, respectively. The energies (\ref{eq-omega}) and the
normalized solutions (\ref{eta1}) are doubly periodic in
$\alpha$ with the periods $2K, 2iK'$.

The solutions (\ref{eta1}) are quasi-periodic in $2K$, and
        \begin{eqnarray}
        \eta(z+2K, \alpha) &=&
         -\eta(z,\alpha) \exp(2K {\rm Z}(\alpha))
         \label{quasiper}
        \end{eqnarray}
Introducing the Floquet index $\nu$ by $e^{-2i\nu K} = - e^{2K
{\rm Z}(\alpha)}$, we  have
        \begin{equation}
        \nu \equiv i {\rm Z}(\alpha)+\frac{\pi}{2K}
        =\frac{\pi n}{QK}
        \label{allowed-nu0}
        \end{equation}
with integer $n$. Returning to the original variable $x=  z k/m$,
 we find the
allowable wave vectors  in the form
        \begin{equation}
        q =2 \pi\bar\rho (i (K/\pi){\rm Z}(\alpha)+ 1/2)
        ={2\pi n}/{L}
        \label{allowed-nu}
        \end{equation}
Consider first the variation of the index $\alpha$
on the segment I, the corresponding $n$ in
(\ref{allowed-nu}) in the range
        \begin{equation}
        -Q/2 < n \leq Q/2,
        \label{allowed-n2}
        \end{equation}
which means $Q$ allowed values of $n$.
The energies (\ref{eq-omega}) of solutions $\eta(z,\alpha)$,
with $\alpha$ from the segment I,
lie between
\[\omega=0, \quad q=0  \quad{\rm at}\quad  \alpha=K\pm iK', \]
and
     \begin{equation}
     \omega =\omega_1\equiv mk_1/k, \quad q=\pi\bar\rho
     \quad{\rm at}\quad
     \alpha=K.
     \label{omega1}
     \end{equation}

\noindent
This part of the spectrum is
interpreted as a band, originated due to hybridization of the
 bound states
related to individual kinks.

In order to see this, we note that for one kink, $Q=1$, the only
value $n=0$ is allowed in the segment I.  It is the quantized
bound state with $\omega=0$ for the single-kink solution of
the sine-Gordon equation.  In the limit of low density
$\bar\rho\ll m$, the energies of the lower band are
exponentially small,
        \begin{equation}
        \omega \simeq 4 m
        e^{-m/2\bar\rho} \sin\frac {|q|}{2\bar\rho},
        \label{disp-k=1}
        \end{equation}
which is the dispersion for a system of weakly coupled harmonic
oscillators.

Another branch of the spectrum, parametrized by $\alpha$ from
the segment II, has one singular
point at $\alpha= -iK'$, where both ${\rm Z}(\alpha)$ and
${\rm dn}(\alpha)$ have simple poles. This point corresponds to
$q\to \pm\infty$ and $\omega\to\infty$.

The energy of this band
has the minima
     \begin{equation}
     \omega=\omega_2 \equiv m/k \quad{\rm at}\quad
 |q|=\pi\bar\rho
     ,\quad \alpha=-2iK',0.
     \label{omega2}
     \end{equation}

\begin{figure}
\includegraphics[width=7cm]{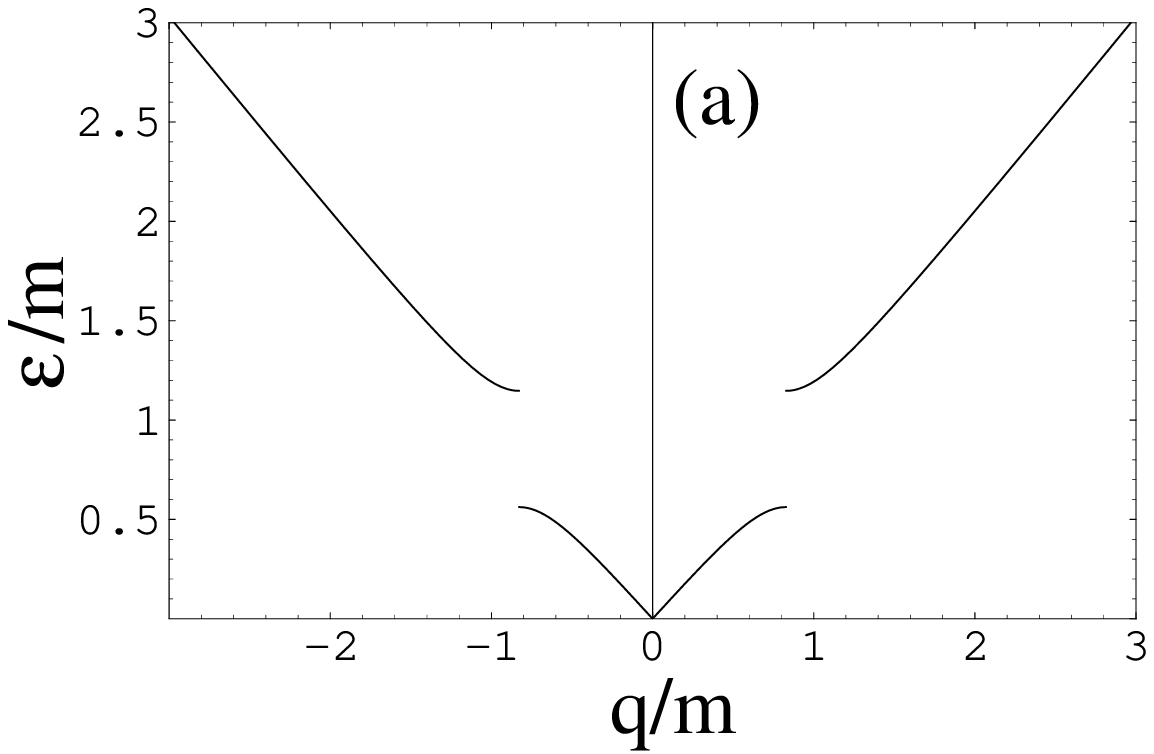}
\vskip1mm
\includegraphics[width=6cm]{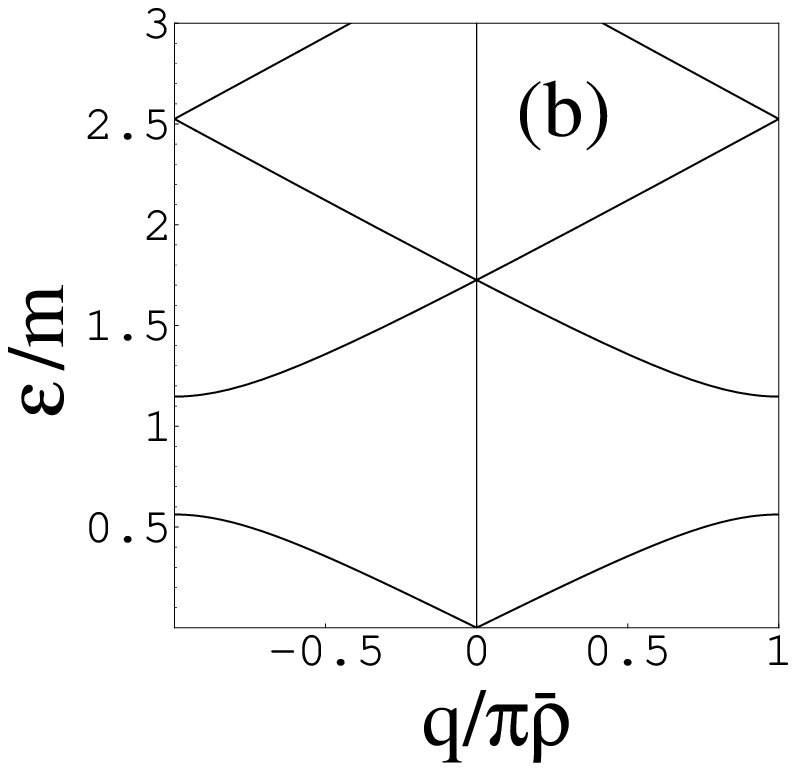}%
\caption{The dispersion in the (a) extended and (b) reduced
 Brillouin zone scheme
shown at $k=0.87$ and  $\bar\rho/m= 0.26$.}
\label{fig:disp}
\end{figure}

The dispersion at $\bar\rho/m= 0.26$ is shown in Fig.\
 \ref{fig:disp}a in the
extended Brillouin zone scheme. The same dispersion shown in the
 reduced
Brillouin zone scheme is depicted in Fig.\ \ref{fig:disp}b.

We see that there is an energy gap $\omega_2-\omega_1$ between
the states in the lower and upper band. It is interesting to
note a following relation \[ \omega_2^2-\omega_1^2 = m^2. \]

In the low-density limit, $\bar\rho\ll m$, we write $\alpha =
i\alpha'$, then $q\to - m\tan \alpha'$, $\omega \to
m/\cos\alpha'$ so that $ \omega = \sqrt{m^2+q^2}$.

At this point it is worthwhile to calculate the Fermi velocity
 $v_F$ for
low-lying excitations in the large-$Q$ limit. Expanding $\omega$
and $q$ near the point $\alpha_0 = K-iK'$ of sector I, we have
        \begin{equation}
        \omega \simeq -i m \frac{k_1}{k}
        (\alpha-\alpha_0), \quad
        q \simeq -i 2\bar\rho E (\alpha-\alpha_0),
        \end{equation}
and
        \begin{eqnarray}
         v_F &=& \omega/q = k_1 K/E ,
        \label{v_F}
        \\
        &\simeq&
       \sqrt{(h-1)\ln \left(\frac{8}{h-1}\right) }
        , \quad h\simeq 1,
         \\
        &\simeq&
        1 -\left(\frac\pi{2h}\right)^4 , \quad h\gg 1,
        \end{eqnarray}
in agreement with the above estimate and the results in
\cite{PapaTsve}.  As a result, the Luttinger parameter, ${\cal
K} =\pi \chi v_F$, is found in the form :
        \begin{eqnarray}
        {\cal K} &=& \frac{\pi\beta^2}{16}
        \frac1{k_1 K^2},
        \label{LuttParam}
        \end{eqnarray}
which shows that $\cal K$ decreases with increasing the density,
 and attains
its limiting value $\beta^2/(4\pi)$ at $\bar\rho \to \infty$.

The form of the eigenfunctions (\ref{eta1}) is somewhat
simplified at the special points $\alpha=K-iK',K,0$,
where one has (cf.\  (\ref{omega1}), (\ref{omega2}))

     \begin{eqnarray}
     \eta_{0}(x) &=& (L E/K)^{-1/2} {\rm dn}(xm/k),
     \quad  \alpha=K-iK'
     \label{zeromode}
     \\
     \eta_{1}(x) &\propto&  {\rm cn}(xm/k),
     \quad  \alpha=K
     \\
     \eta_{2}(x) &\propto&  {\rm sn}(xm/k),
     \quad  \alpha=0
     \end{eqnarray}
with the normalization factor in $\eta_{0}(x)$ made
explicit.  Henceforth, we will call
the eigenfunction (\ref{zeromode}) the zero mode.

Analyzing the lowest-energy states with $\alpha \simeq K-iK'$,
one can find a following expression,
valid in the leading order in $q/\pi\bar\rho$ :

        \begin{eqnarray}
        \eta_{q}(x) &=& \eta_{0}(x)
        \exp\left(-iqx
         - i \frac{q\pi}{4KE\bar\rho}\,
        \frac{\vartheta'_3(\pi x\bar\rho)}
        {\vartheta_3(\pi x\bar\rho)}
        \right).
        \label{eta-smallq}
        \end{eqnarray}
The second term in the exponential is the phase of the Bloch
 function.
It is $1/\bar\rho-$periodic  in $x$, vanishing at
$x=0, \pm\bar\rho^{-1}, \pm2\bar\rho^{-1}, \ldots$.

For completeness, we provide here an approximate expression for
the eigenfunctions of the upper band in the low-density limit,
$\bar\rho \ll m$ and $k\to1$. Near the position of $n$th
kink, it is given by
       \begin{eqnarray}
       \eta(x)_{\omega\simeq m} &\propto&
       ({\rm tanh}\, m(x-\bar\rho n) -i\tan\alpha)
       \nonumber \\ &&\times
       \exp({i(x-\bar\rho n)m\tan\alpha -in q/\bar\rho }),
       \label{higher-eta}
       \\
       q &=& -m\tan\alpha+2\alpha\bar\rho+\pi\bar\rho,
       \end{eqnarray}
in accordance with \cite{MorFes}.

Concluding this subsection, we note that eqs.\ (\ref{v_F}),
(\ref{LuttParam}), (\ref{eta-smallq}), (\ref{higher-eta}) extend
the analysis of fluctuations performed by previous authors.
\cite{Lebwohl,Kutnjak,Rodriguez}

\subsection{Quantization of Fluctuations}

The differential operator in (\ref{eq:Schro})
is self-adjoint and, consequently, the eigenfunctions
form a orthonormal set,
        \begin{equation}
        \int_0^L dx \eta_{n}(x) \eta_{m}(x) = \delta_{nm}
        \end{equation}

\noindent
There is a well-known property that a continuous function $F(x)$
(with the same boundary conditions) can be expanded as a
 generalized
Fourier series
        \begin{equation}
        F(x) = \sum_{n}
        \eta_{n}(x)
        \int_0^L dy \eta_{n}(y) F(y)
        \label{genFou}
        \end{equation}
which means
        \begin{equation}
        \sum_{n}
        \eta_{n}(x)
        \eta_{n}(y) =\delta(x-y)
        \label{deltafun}
        \end{equation}

Now we may introduce the quantized field, defining
        \begin{equation}
        \Phi(x,t) = \sum_{n}
        \frac{1 }{\sqrt{2\omega_{n}}}
        \left(
        \eta_{n}(x) e^{i\omega_{n}t} b_{n}^\dagger
        + \eta_{n}^\ast(x) e^{-i\omega_{n}t} b_{n}
        \right)
        \label{qfield}
        \end{equation}

\noindent
The Bose operators $b_n$ satisfy the commutation relations
        \begin{equation}
        [b_{n}, b_{m}^\dagger] = \delta_{{n},{m}}
        \end{equation}
and therefore for equal-time fields
        \begin{eqnarray}
        [\Phi(x), \partial_t \Phi(y)] %|_{t=0}
        &=&
        \frac{i}2 \sum_{n}
        \left(
        \eta_{n}(x) \eta_{n}^\ast(y)
        + \eta_{n}^\ast(x) \eta_{n}(y)
        \right)
        \label{commut} \\ &=&
        i \delta(x-y)
        \label{deltafun1}
        \end{eqnarray}
Strictly speaking, the zero mode (\ref{zeromode}) with
$\omega=0$ should be excluded from the sum (\ref{qfield}), which
leads to the extra term $-i\eta_0(x)\eta_0(y) \sim Q^{-1}\to 0$
in (\ref{commut}). The reason for it is  non-independent
character of the mode $\eta_0$ in the decription of the field
configuration, $\phi_0(x), \{\eta_n(x)\}_{n=0}^\infty$. Indeed,
$\eta_0$, bearing the zero energy, corresponds to the
translation of the classical solution $\phi_0(x) \to
\phi(x+X)$, while $\eta_0(x)\sim \partial_x\phi_0(x)$. As a
result, the consistent treatment will consist of the functions
$\phi_0(x+X), \{\eta_n(x+X)\}_{n=1}^\infty$, with $\eta_0$
replaced by the displacement $X$ of the system as whole. The
quantity $X$ should be regarded as a dynamical variable in the
correct description of the quantum Hamiltonian. \cite{ChristLee}

We outline this collective coordinate method in Appendix and
show, that the corrections to the above simpler formula are
negligible in the limit $Q\to\infty$. The only thing, which we
should borrow from the collective coordinate method, is the
necessity to integrate over the vacuum location $X$ instead of
inclusion of $\eta_0$.

It seems that one now has everything to compute
various correlation functions, describing the quantum
fluctuations of the system.
Given a set consisting of the classical
solution $\phi_0$ and the normal modes around it, one calculates
the quantum averages with the simple rules of bosonic algebra
and then integrates over the initial position $X$ of the soliton
lattice. This program however reveals many difficulties
described in the next section.

\section{Difficulties in the Approach}
\label{sec:diffic}

\subsection{Density-type correlations}

Consider the fermionic density $\rho = \partial_x \beta
\phi/2\pi$. The classical contribution $\rho_0 = \partial_x \beta
\phi_0/2\pi = \bar\rho w(x)$ where
\begin{equation}
w(x) \equiv \frac{2K}{\pi} {\rm dn}\frac{m(x+X)}{k}
\end{equation}
The averaging over the kinks position gives a simple
result here,
$$\langle \rho(x) \rangle_X = \bar\rho
\langle w(x) \rangle_X = \bar\rho. $$
At the same time, the classical contribution to the pairwise
density correlator is
$\langle \rho(x) \rho(y) \rangle_X = \bar\rho^2
\langle w(x)w(y) \rangle_X \neq \bar\rho^2$.
These non-trivial correlations of density at largest distances
correspond to Wigner crystallization of the
charge $Q$. This long-range ordering of the charge
in one spatial dimension is evidently erroneous and should
eventually be removed from the theory.

The quantum corrections to the above correlator in the
long-distance limit read as
\begin{equation}
\langle \rho(x) \rho(y) \rangle_q = -\frac{{\cal K}}{2\pi^2}
\frac{\partial^2}{\partial x \partial y}
\langle w(x)w(y) \rangle_X
\ln |x-y|
\label{cor-den1}
\end{equation}
where we have used (\ref{v_F}), (\ref{zeromode}),
(\ref{eta-smallq}),  (\ref{qfield}).

The CDW-type correlator is similarly estimated as
\begin{equation}
\langle e^{i\beta \phi(x)} e^{-i\beta\phi(y)} \rangle \sim
\langle \exp(-2{\cal K} \ln (x-y) w(x)w(y)) \rangle_X
\label{cor-CDW1}
\end{equation}
Evidently, the modulation $w(x)w(y)$ in (\ref{cor-CDW1}), being
exponentiated is more pronounced than in (\ref{cor-den1}).

\subsection{Dual Field}

The so-called dual field $\theta(x,t)$
used in the construction of the fermion operator,
is usually defined as
$$
\partial_x \theta = \partial_t \phi, \quad
\partial_t \theta = \partial_x \phi.
$$
These relations imply $\partial_t^2 \phi
 = \partial_x^2 \phi$, which is not consistent with
the linearized theory equation (\ref{Schro}) (the discussion of
the zero mode contribution to $\theta$ is postponed until the
next section).  Particularly, the usual construction of the
charge-raising operator ${\cal O} \sim\exp 2\pi
i\beta^{-1}\int^{(x,t)} (dt \partial_x+dx\partial_t)\phi$ is
inappropriate in the basis $\{\eta_n\}$, being dependent on the
contour of integration in $\cal O$.

One may ignore for a moment this discrepancy, using
only a first equation $\partial_x \theta = \partial_t \phi$.
This is sufficient for the calculation of the instantaneous
correlation function $\langle \exp({4\pi i\beta^{-1}\theta(x)})
\exp({-4\pi i\beta^{-1}\theta(y)}) \rangle$.
Making this calculation, we meet another troublesome property.

The lattice of kinks gives rise to the notion of the Brillouin
zone and to the infinite set $\{ \eta^{(0)}_n \}$ of the
eigenstates with non-zero energy and zero wave vector in
the reduced Brillouin zone scheme; $\eta^{(0)}_0\equiv \eta_0$.
A constant is not an eigenstate of (\ref{Schro}), and therefore
$\bar\eta_n \equiv L^{-1}\int_0^L dx\, \eta^{(0)}_n(x) \sim
L^{-1/2}$.  Further, if the wave vector $q_n$ of the eigenstate
$\eta_n$ is close to zero, a corresponding Fourier transform
$\eta_n(q)$ has a pole $\bar\eta_n/(q-q_n)$.
The quantity $\bar\eta_n$ decreases when increasing the energy
and this decrease is slower at larger $k$.  In particular, one
can show for the above set $\{ \eta^{(0)}_n \}$ that
$\sum_{n=0}^\infty \left(\bar\eta_n)\right)^2 =L^{-1}$,
while $\bar\eta_0 =\pi/(2\sqrt{LKE})$.

The vacuum expectation value
of the product of $\theta$ fields is then
        \begin{eqnarray}
       \langle \theta(x+y) \theta(y) \rangle &\simeq &
         \sum_{q>0} \frac{v_F}{ q}
        \bar\eta_{0}^2 \cos qx
        \nonumber \\ &&
        + \sum_{q'} \frac{\cos q'x}{2 q'^2}
        \sum_{n\geq 1}\omega_{n,q'} \bar\eta_{n}^2 ,
        \label{correl-theta}
        \end{eqnarray}
As usual, the constant (divergent) term should be added here in
 order
to analyze the dependence of the correlations on $x$.
The first term in (\ref{correl-theta}), resulted from the
lowest energy modes, contributes to the logarithm.
In the second term, with both signs of $q'$ available,
we note that linear-in-$q'$ corrections to $\omega_{n,q'}$ vanish
 and
the sum over $q'$ gives a contribution linear in $|x|$.
The sum over $n$ in (\ref{correl-theta})
converges rapidly when elliptic index $k$ is not too
close to unity, i.e. at $\bar\rho \agt m$. For an estimate, it is
 sufficient to
write $\sum_{n\geq 1}\omega_{n,q'} \bar\eta_{n}^2 \sim
\bar\rho\sum_{n\geq 1}\bar\eta_{n}^2 = \bar\rho
 (L^{-1}-\bar\eta_0^2)$.
As a result, we have
        \begin{eqnarray}
        \langle \theta(x+y) \theta(y) \rangle &\simeq &
        \frac{\pi k_1}{8E^2} \ln \frac{L}{|x|}
        \nonumber \\ &&
        - \bar\rho |x| (1-\pi^2/(4KE)) O(1)
        \end{eqnarray}
The linear in $|x|$ term is present unless $k\to 0 $ ($m/\bar\rho
 \to0$), when
one recovers the usual expression,
$\langle \theta(x+y)  \theta(y) \rangle \sim -\ln |x|$. As a
 result,
the superconducting correlations
$\langle e^{4\pi i\beta^{-1}\theta} e^{-4\pi
i\beta^{-1}\theta}\rangle$ decay exponentially at large
distances.

We see that the above definitions of the
dual field are inconsistent in the basis $\eta_n$ and lead to
an exponential decrease of the SC correlations in the gapless
situation.

\subsection{Perturbation Theory}

\begin{figure}[ht]
\includegraphics[width=7cm]{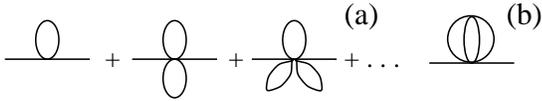}% \vskip1mm
\caption{(a) A tadpole sequence of diagrams, contributing to
position-dependent mass renormalization. (b) A stronger divergent
diagram not contained in this sequence.}
\label{fig:massren}
\end{figure}

Consider now the role of the terms of the interaction
in (\ref{L-expa}), which have been discarded.
These terms, containing additional $\beta$s,
produce divergences in the perturbation theory. Resumming the
tadpole sequence of diagrams, Fig.\ \ref{fig:massren}a, one
 obtains the mass
renormalization of the form
\[
\eta^2(x) m^2 \cos\beta\phi_0(x) \to \eta^2(x) m^2
\cos\phi_0(x) e^{-{\cal K} w(x)^2 \ln \Lambda} \]
with $\Lambda \sim Q$. This renormalization depends on $x$ and,
therefore, changes the shape of the potential $\cos\beta\phi_0$.
This essential modification is accompanied by stronger
than logarithmic infra-red divergencies in
diagrams, not contained in the tadpole series, e.g., that shown
in Fig.\ \ref{fig:massren}b.

\section{a proposed solution}
\label{sec:solu}

\subsection{Lagrangian}

A key to resolve all the above discrepancies is found as
follows.
We wrote the fluctuations around the classical solution as
$\phi=\phi_0+\eta$ and performed the Taylor expansion
(\ref{L-expa}).
Instead of that, we can write the fluctuations in a form
        \begin{equation}
        \phi = 2\beta^{-1} {\rm am}\left[
        2K \left(x\bar\rho +
        \frac{\beta\tilde\eta}{2\pi}\right)  \right].
        \label{newdefphi}
        \end{equation}
The formula (\ref{newdefphi}) has a remarkable property, that
 shifting
$\tilde\eta \to \tilde\eta+2\pi/\beta$, we increase accordingly
$\phi \to \phi +2\pi/\beta$. This may serve as a justification of
 the
scaling factor before $\tilde\eta$ in (\ref{newdefphi}).
At $\bar\rho\gg m$ one has
$\phi\to 2\pi\bar\rho\beta^{-1} x + \tilde\eta$,
hence (\ref{newdefphi}) and the previous writing 
$\phi=\phi_0+\eta$ coincide in this limiting case.  Some 
calculation shows, that the perturbation theory in $\tilde\eta$ 
can be obtained from the exact form of the full Lagrangian
  \begin{eqnarray}
   {\cal L}&=&
 w^2\left[x+\frac{\beta\tilde\eta}{2\pi\bar\rho} \right]
 \frac{
  (\partial_t\tilde\eta)^2  -
 (\partial_x \tilde\eta)^2 }2.
 \label{lagr-best}
 \end{eqnarray}
The field theory decribed by (\ref{lagr-best}) is free from
infra-red divergencies for two reasons. The first  is the
presence of the field derivatives in each term of the
perturbation series. Another reason is a particular form of the
formfactors of the interaction, when $n$th vertex is obtained
by $(n-2)$ differentiations of the restricted function $w^2(x)$.

The Gaussian action is given by
 \begin{eqnarray}
 {\cal L}&=& \frac12 w^2(x)\left( (\partial_t\tilde\eta)^2  -
 (\partial_x \tilde\eta)^2  \right).
 \label{lagr-mod}
 \end{eqnarray}
with the corresponding equation of motion
 \begin{eqnarray}
 \partial_t^2 \tilde\eta &=& w^{-2}\partial_x
 w^2 \partial_x  \tilde\eta
 \label{nueqmot}
 \end{eqnarray}

Our aim is now to show that the spectrum of (\ref{nueqmot})
coincides with one of the initial problem (\ref{Schro}), and to
find the connection between the fields $\tilde\eta$ and $\eta$.

The equivalence of the spectra stems from the possibility
to rewrite the Lam\'e equation in the form :
 \begin{eqnarray}
 \omega^2 \eta(x) &=& -w^{-1}(x)\partial_x
 w^2(x) \partial_x
 w^{-1}(x) \eta(x)
 \label{A+A}
 \end{eqnarray}
which implies that the Lagrangian is
 \begin{eqnarray}
 {\cal L}&=&\frac12\left( (\partial_t\eta)^2  -
 (w \partial_x w^{-1}\eta)^2 \right)
 \label{lagr-1}
 \end{eqnarray}
Evidently, eq.\ (\ref{lagr-1}) coincides with (\ref{lagr-mod})
after the substitution
  \begin{equation}
  \tilde\eta(x,t) \equiv   w^{-1}(x) \eta(x,t).
  \label{2etas}
  \end{equation}
As a result, the eigenfunctions to (\ref{lagr-mod}) are
normalized with a weight
      \begin{equation}
      \int dx\, w^2(x)
      \tilde\eta_\alpha(x) \tilde\eta_\beta(x)
      =\delta_{\alpha\beta},
      \end{equation}
and (\ref{2etas}) leads to the following form of the
correlations at large distances
   \begin{equation}
   \beta^2 \langle \tilde\eta(x,t)\tilde\eta(y,0) \rangle
   \simeq  -{\cal K} \ln|(x-y)^2-v_F^2t^2|
   \end{equation}

Therefore, we find the description of the
fluctuations, possessing the same spectrum in the Gaussian
approximation, but free from divergencies in higher orders of
interaction. This choice of appropriate variables should
be hence considered as an effective partial resummation of
perturbation series.

\subsection{Density-density correlations}

The adoption of the definition
(\ref{newdefphi}) makes the calculation of the correlation
functions less trivial.  Consider first the density-density
correlations $\langle \rho(x) \rho(y) \rangle$. Using the series
representation \cite{Ab-St} for the function ${\rm dn}(x)$ we
 write

  \begin{equation}
  \rho =
  \left(
  \bar\rho +\frac{\beta}{2\pi}\partial_x\tilde\eta
  \right)
  \sum_{n=-\infty}^\infty
  \frac{e^{in(2\pi\bar\rho x+\beta\tilde\eta)}}
  {{\rm cosh} n\tau},
  \end{equation}
with $\tau=\pi K'/K$. Note that the previous
kinks' lattice displacement $X$ corresponds to the new zero
mode $\tilde\eta_0=const$, eq.(\ref{2etas}). The integration
over $\tilde\eta_0$ gives evidently $\langle \rho \rangle
 =\bar\rho$,
while it selects the terms with equal $n$s in the pairwise
 density
correlator,
    \begin{eqnarray}
    \langle \rho(x) \rho(y) \rangle_q &=&
    \bar\rho^2 + \frac{{\cal K}}{2\pi^2(x-y)^2}
    \label{cor-den2}
    \\
    &+&
    \sum_{n=1}^\infty c_n
    \frac{\cos 2n \pi\bar\rho (x-y) }
    {|x-y|^{2n^2{\cal K}}} + \ldots
    \end{eqnarray}
with omitted faster decaying terms. We see that
the weight factors $w(x)$ in (\ref{cor-den1}) are transformed
into the decaying $4k_F=2\pi\bar\rho$ harmonics of the density,
in line with Ref.\ \cite{haldane82}.

\subsection{
CDW-type correlations and
umklapp ordering}

Keeping in mind the Hubbard model near half-filling, we
associate the exponent $e^{i\beta\phi/2}$ with the ``$2k_F$''
density wave operator, whereas $e^{i\beta\phi}$
stands for the ``$4k_F$'' umklapp process.

The ``$2k_F$'' correlations are estimated as follows. We use
(\ref{newdefphi}) and the series representation
    \begin{eqnarray}
    e^{i {\rm am} (2Kz)}
    &=&
    \sum_{n=-\infty}^\infty
     C^{(1)}_n   e^{2\pi iz(n+1/2)},
    \label{eiam-series}
    \\
    C^{(1)}_n  &=&
    \frac{\pi e^{\tau(n+1/2)}}
    {k K {\rm sinh} (2n+1)\tau},
     \nonumber
    \end{eqnarray}
to obtain

    \begin{eqnarray}
    \langle e^{i\frac\beta2 \phi(x)} e^{-i\frac\beta2\phi(0)} 
    \rangle
    &\sim&
    \sum_{n=-\infty}^\infty
    [C^{(1)}_n]^2
    \frac{e^{2i\pi \bar\rho x(n+1/2)}}
    { |x|^{2{\cal K}(n+1/2)^2 } }
    \label{cor-CDW2a}
    \\
    \langle e^{i\frac\beta2 \phi(x)} e^{i\frac\beta2\phi(0)} 
    \rangle
    &\sim&
    \sum_{n=-\infty}^\infty
    C^{(1)}_n C^{(1)}_{-1-n}
    \frac{e^{2i\pi \bar\rho x(n+1/2)}}
    { |x|^{2{\cal K}(n+1/2)^2 } }
    \label{cor-CDW2b}
    \end{eqnarray}

The prefactors of the exponents here are not universal,
depending on the  cutoff. What matters here, is the ratio
of the amplitudes for the forward-going and
backward-going waves with the same absolute value of $(n+1/2)$.
The leading asymptotes are defined by $n=0,-1$ ;
noting that $C^{(1)}_{-1}=-\tilde q C^{(1)}_{0}$, with
$\tilde q=e^{-\pi K'/K}$, we write
    \begin{eqnarray}
    \langle e^{i\beta \phi(x)/2} e^{-i\beta\phi(0)/2} \rangle
    &\sim&
     ( e^{i\pi \bar\rho x} +\tilde q^2 e^{-i\pi \bar\rho x})
     |x|^{-{\cal K}/2 } .
    \\
    \langle e^{i\beta \phi(x)/2} e^{i\beta\phi(0)/2} \rangle
    &\sim&
     - 2\tilde q \cos({\pi \bar\rho x})
     |x|^{-{\cal K}/2 } .
    \end{eqnarray}
Particularly, the above formulas indicate the following.
The lowest harmonic of the CDW
order parameter field,
$\propto \sin(\beta \phi/2) = {\rm sn}
[2K(\bar\rho x +\beta\tilde\eta/2\pi)]$
has an amplitude  $\pi/(kK\sinh\tau/2)$ which
remains finite in the total range of
variation og $\bar\rho$. The similar quantity for the SDW field,
$\propto \cos(\beta \phi/2) = {\rm cn} [2K(\bar\rho x
+\beta\tilde \eta/2\pi)]$, has an amplitude  
$\pi/(kK\cosh\tau/2)$ which vanishes as $\bar\rho/m$ at
the half-filling, $\bar\rho=0$.

Using (\ref{eiam-series}) we write the umklapp order parameter
field in the form
    \begin{eqnarray}
    \cos{\beta\phi}
    &=&
    \sum_{n=-\infty}^\infty
    C^{(2)}_n e^{in(2\pi\bar\rho x + \beta\tilde\eta)},
    \\
    2C^{(2)}_n &=&
    \sum_{n_1}  C^{(1)}_{n-n_1} C^{(1)}_{n_1}
    +(n\to -n).
    \label{defcoefC}
    \end{eqnarray}
We observe the appearance of the constant term, $C^{(2)}_0$,
corresponding to the fact that the average value $\langle
e^{i\beta \phi(x)}  \rangle = \langle \cos{\beta \phi(x)}
\rangle = C^{(2)}_0$ is not zero at  finite density of kinks.
Indeed one has
     \begin{eqnarray}
    C^{(2)}_0 &=& (-2E+K(1+k_1^2))/(k^2K) , \\
     &\simeq&  1- 4 \bar\rho/m , \quad \bar\rho /m \alt 0.2 \\
    &\simeq& m^2/(8\pi^2 \bar\rho^2) , \quad \bar\rho /m \agt 0.2
    \label{umklapp}
    \end{eqnarray}
It should be noted that the non-zero value of $C^{(2)}_0$
arises without ``locking'' of the field $\phi$. We
remind that in the absence of kinks the field $\phi$ is
``locked'' to one of the discrete values $\phi_n = 2\pi n/\beta$,
and it results in the non-zero value of the ``$2k_F$'' density
wave $\langle e^{i\beta\phi/2} \rangle =\pm1$. In the presence
of kinks $\phi$ connects different vacua $\phi_n$, and
$\langle e^{i\beta\phi/2} \rangle = 0$, while $\langle
e^{i\beta \phi}  \rangle \neq 0$ is decreasing function of
the kinks density.

The quantum average of
two umklapp exponents at long distances has a form
    \begin{eqnarray}
    \langle \cos{\beta \phi(x)} \cos{\beta\phi(0)} \rangle
    &\sim& [ C^{(2)}_0 ]^2 +
    \sum_{n=1}^\infty
    c_n' % [C^{(2)}_n]^2 
    \frac{\cos{2\pi \bar\rho x n)}}
    {|x|^{2{\cal K}n^2}}
    \label{cor-umklapp}
    \end{eqnarray}
We see that in the absence of $C^{(2)}_0$
the dimensionality of the operator $\cos{\beta \phi }$
would be $\cal K$. 
As $\cal K$ increases with decrease of $\bar\rho$,
one might think that umklapp operator $\cos{\beta \phi }$
becomes less relevant. \cite{PapaTsve} 
However, this increase of $\cal K$ is
accompanied by the increase of $C^{(2)}_0$, and
$\langle \cos{\beta \phi } \rangle \sim 1$ at $\bar\rho\sim m$.

\subsection{Dual field}

Let us now discuss the modification of the definition of the
dual field $\theta(x,t)$.
The basic property of the charge raising operator
${\cal O} \sim\exp 2\pi i\beta^{-1} \theta$ in the
path integral representation is
that the contour of integration  (``Dirac string'') in the
 definition of
$\theta$  introduces a discontinuity in the field $\phi$, so that
 the values
of the field $\beta\phi$ differ by $2\pi$  across this contour.
\cite{Lukyanov01}
In view of the above property,
$\beta \phi \to \beta\phi +2\pi$ as $\beta\tilde\eta \to
\beta\tilde\eta+2\pi$, we see that it suffices to require
the charge-raising property for the field $\tilde\eta$.

The canonical momentum for the field $\tilde\eta$
is given by $\tilde \pi = \partial {\cal
L}/\partial( \partial_t \tilde\eta) = w^2\partial_t \tilde\eta$.
The quantization condition reads as $[\tilde\eta(x),\tilde\pi(y)]
 =
i\delta(x-y)$.

Consider the definition
    \begin{equation}
    \tilde\theta(x) =\int^x dx'\,\tilde\pi(x')
    \label{newDual1}
    \end{equation}
for equal-time fields.
Differentiating the last equality by $x$ and then by $t$ we have
from (\ref{nueqmot})
$\partial_t\partial_x \tilde\theta = \partial_x w^2 \partial_x
\tilde\eta$, hence the definitions
    \begin{equation}
    \partial_x\tilde\theta  =w^2 \partial_t \tilde\eta,
    \quad
    \partial_t\tilde\theta  =w^2 \partial_x \tilde\eta,
    \label{newDual}
    \end{equation}
are consistent with the equation of motion. The charge-raising
operator,  ${\cal O} \sim\exp 2\pi i \beta^{-1}\int^{(x,t)}
w^2(dt \partial_x+dx\partial_t)\tilde\eta$, is now uniquely
defined, independent of the contour of integration.  Further,
one has $[\tilde\eta(x), \tilde\theta(y)] = i\vartheta(y-x)$ and
therefore  we have
$\exp(2\pi i\beta^{-1}\tilde\theta (x)) f(\tilde \eta(y)) =
f(\tilde \eta(y) + 2\pi\beta^{-1}\vartheta(x-y))
\exp(2\pi i\beta^{-1}\tilde\theta (x))$
for any $f(z)$. This property allows one to define the fermion
operator, see below.

A counterpart $\tilde\theta_0$ of $\tilde\eta_0 =const$ cannot
be obtained from (\ref{newDual}), which shows again a particular
role of the zero mode. One of the possible ways to introduce the
quantized quantity $\tilde\theta_0$, relevant to our discussion
in Appendix, is found in \cite{Banks}.  One can write these
zero-energy components as follows ~:
    \begin{eqnarray}
    \tilde\eta_0 &=&
    \frac{\pi^2\beta }{8LEK} J t
    +(q-Q) W_1(x) \frac{\beta}{2\pi\chi L}
    +i\frac2\beta \frac{\partial}{\partial J}
    \label{zeroDualA} \\
    \tilde\theta_0 &=&
    \frac{\pi^2\beta }{8LEK}
    J W_2(x) +(q-Q) t \frac{\beta}{2\pi\chi L}
    +i\frac{\beta}{2\pi} \frac{\partial}{\partial q}
    \label{zeroDualB}
    \end{eqnarray}
with $J=P/\pi \bar\rho$ the persistent current attaining integer
values, the functions $W_1(x) =\int^x w^{-2}(y) dy$ and $W_2(x)
=\int^x w^{2}(y) dy$ are expressed through the elliptic
functions ${\rm Nd}(x)$ and ${\rm Dn}(x)$, respectively.
A canonical conjugate to $J$ is $\pi\bar\rho X$, so the last
term in (\ref{zeroDualA}) is simply the previous shift in the
initial kink position.
 The charge variable $q$ was assumed to be $Q$ in Sections
\ref{sec:fluctu}, \ref{sec:diffic} and we let it fluctuate now
around its equilibrium value. A choice of the factors before $J$
and $q$ variables is dictated by the condition $\int_0^Ldx \,w^2
[(\partial_t\tilde\eta_0)^2 +(\partial_x\tilde\eta_0)^2 =
(q-Q)^2 \chi^{-1}L^{-1} + P^2 {\cal M}_{tot}^{-1} $, cf.
(\ref{ene-rho}). A particular form of the function $W_1(x)$ in
(\ref{zeroDualA}) is not coincidental here. Satisfying the
equation $ w^{-2}\partial_x w^2 \partial_x W_1(x) =0$, the
function $W_1$ is another (increasing) zero mode for eq.\
(\ref{nueqmot}), which corresponds to the increase of $Q$ in
the classical solution $ w(x)W_1(x) \propto \partial \phi_0
/\partial k$ ; we have $\beta \tilde\eta_0 \left.\right|_{x=0}^L
= 2\pi(q-Q)$.

Let us discuss now the correlations of the dual field
$\tilde\theta$.  To do it, consider first its equation of motion,
as follows from (\ref{newDual}) :
  \begin{equation}
  \partial_t^2 \tilde\theta =
  w^2\partial_x w^{-2}\partial_x \tilde\theta
  \label{eqmotDual}
  \end{equation}
Recalling the property of the Jacobi function
${\rm dn}(x)$, we write
  \begin{equation}
  w(x+\bar\rho^{-1}/2) = c_1 w^{-1}(x),
  \quad c_1= \frac{4K^2k_1}{\pi^2}.
  \label{coef-c1}
  \end{equation}
Therefore the differential operator in the
r.h.s. of (\ref{eqmotDual}) coincides with one in
(\ref{nueqmot}), when shifted by its half-period, $x\to x+
\bar\rho^{-1}/2$. 

For the infra-red Gaussian action,
given some particular solution $\tilde\eta_\alpha$ to
(\ref{nueqmot}), we obtain its dual
counterpart $\tilde\theta_\alpha$ with the same energy and
Floquet index.  In view of the completeness of the set
$\tilde\eta_\alpha$, we have a relation $\tilde\theta_\alpha(x)
= c_\alpha \tilde\eta_\alpha(x+\bar\rho^{-1}/2)$. The value of
 the
constant $c_\alpha$ is elucidated from the following sequence of
equalities
 \begin{eqnarray}
 c_\alpha \partial_t \tilde\eta_\alpha(x+\bar\rho^{-1}/2) &=&
 \partial_t \tilde\theta_\alpha(x) =
 w^2(x) \partial_x \tilde\eta_\alpha(x)
 \nonumber \\ & =&
 e^{2i\nu K}w^2(x) \partial_x \tilde\eta_\alpha(x+\bar\rho^{-1})
  \\ & =&
 e^{2i\nu K}w^2(x) c_\alpha^{-1} \partial_x
 \tilde\theta_\alpha(x+\bar\rho^{-1}/2)
 \nonumber\\ & =&
 e^{2i\nu K}c_1^2 c_\alpha^{-1} \partial_t
 \tilde\eta_\alpha(x+\bar\rho^{-1}/2)
 \nonumber
 \end{eqnarray}
hence
  \begin{equation}
  c_\alpha = \pm c_1 e^{i\nu K}
  \label{coef-ca}
  \end{equation}
One can show, that the sign here is plus for
$\alpha\in (K-iK',K)$ and $\alpha\in (-iK',0)$ and minus for
$\alpha\in (K-2iK',K-iK')$ and $\alpha\in (-2iK',-iK')$.
In the large-$\bar\rho$ limit the eigenfunctions
$\tilde\eta_\alpha(x)$ are usual plane waves, $\sim
e^{-iq_\alpha x}$, and we recover the usual expression,
$\tilde\theta_\alpha(x) = sign(q_\alpha)\tilde\eta_\alpha(x)$.
The sign in (\ref{coef-ca}) is irrelevant for calculating
the expectation value of  the product of two dual fields and we
have

     \begin{equation}
     \langle \tilde\theta(x) \tilde\theta(y) \rangle =
     c_1^2 \langle \tilde\eta(x) \tilde\eta(y) \rangle .
     \end{equation}
Noting from (\ref{LuttParam}),
(\ref{coef-c1}), that $4\pi c_1=\beta^2/{\cal K}$,
we immediately obtain
     \begin{equation}
     \langle e^{4\pi i\beta^{-1}\tilde\theta(x)} e^{-4\pi
     i\beta^{-1}\tilde\theta(y)}\rangle \sim
     |x-y|^{-2/{\cal K}}.
     \label{SC-corr}
     \end{equation}

The dynamics of the correlations (\ref{cor-den2}),
(\ref{cor-CDW2a}), (\ref{SC-corr}) is obtained by replacing
$|x-y|\to ((x-y)^2-v_F^2 (t_1-t_2)^2)^{1/2}$.  This simple
temporal dependence of the correlations is valid as long as we
consider the long-time, long-distance behavior, $x \sim v_F t
\gg \bar\rho^{-1}$.  At shorter scales, the complicated
structure of the wave functions and of the dispersion comes into
play and the Lorentz form of the correlations is lost.

\subsection{Fermionic correlations}

We now discuss the Hubbard model
and regard $\cos \beta\phi$ as four-fermion umklapp  term.
In this case, in addition to the ``charge'' field $\phi$
discussed insofar, one also considers the spin degree of
freedom described by the field $\phi_s$ and its dual one
$\theta_s$. We write the right and left fields as the linear
combinations

       \begin{eqnarray}
       \phi_{R\sigma} &=&
       \frac\beta4 \phi - \frac{2\pi}\beta \tilde\theta
       +\sigma \sqrt{\frac\pi2} (\phi_s -\theta_s )
       \label{Rmover}
       \\
       \phi_{L\sigma} &=&
       \frac\beta4 \phi + \frac{2\pi}\beta \tilde\theta
       +\sigma \sqrt{\frac\pi2} (\phi_s +\theta_s )
       \end{eqnarray}
with the spin projection $\sigma=\pm1$.
The operators
        \begin{equation}
        \psi_{R\sigma} = \chi_\sigma  e^{i\phi_{R\sigma}}
        \quad
        \psi_{L\sigma} = \chi_\sigma  e^{-i\phi_{L\sigma}}
        \label{fermion}
        \end{equation}
may be identified with the right
and left fermion, respectively. Let us discuss this point in
more detail.

One can easily check, that given the usual relation
$[\phi_s(x),\theta_s(y)]=i\vartheta(y-x)$, the operators
$\psi^\dagger_{R\sigma}, \psi_{R\sigma}, \psi^\dagger_{L\sigma},
\psi_{L\sigma}$ with the same $\sigma$ anticommute. The
anticommutation of these operators with different projection of
spin is ensured by the Majorana fermionic variables
$\chi_\sigma$. \cite{Banks} These variables obey
$\chi_+\chi_+ = \chi_-\chi_-=1$, while
$\chi_+\chi_- = -\chi_-\chi_+$.

A subtler question concerns the relation of the construction
(\ref{fermion}) to the initial fermionic variables in the
Hubbard model.
In the bosonization procedure, the initial fermions were
written similarly to (\ref{fermion}), with the dual field
$\theta$ in (\ref{Rmover}) instead of our $\tilde\theta$. 
The
oscillator representation (\ref{qfield}) for the initial
right(left) bosons had simple form, which is achieved 
within our formalism for $\bar\rho\neq 0$ at $m=0$. 
It is not at once clear that the complicated expressions
(\ref{newdefphi}), (\ref{newDual}), (\ref{eta1}) in the presence
of umklapp term, $m\neq 0$ are related to right and 
left moving particles.  
We state here that the label ``right'' of the boson field 
$\phi_{R\sigma}$ simply indicates the species of fermion, which 
is constructed this way.  As we will see shortly, umklapp 
interaction produces a partial redistribution of the fermionic 
spectral weight over the $4k_F$-components of the density, hence 
initially right-moving particle has now left-going components.

In short, our fermions (\ref{fermion}) anticommute, have the same
initial field $\phi$, and explicitly acquire the conventional 
form at $m=0$. Therefore they are the fermions, which one 
starts with during the bosonization of the Hubbard model.

The spin fields of the fermions are factorized, and for the 
charge part we have
       \begin{eqnarray}
        \psi_{R,L}(x)
       &\propto&
       \exp(\pm i\frac\beta4 \phi - i\frac{2\pi}\beta
       \tilde\theta)
       \\
       &\sim&
       e^{- 2\pi i\beta^{-1} \tilde\theta}
       \sum_{n=-\infty}^\infty
       C^{(h)}_n
       e^{\pm i(2\pi\bar\rho x + \beta\tilde\eta)(n+1/4)},
       \\
       C^{(1)}_n &=&
       \sum_{n_1}  C^{(h)}_{n-n_1} C^{(h)}_{n_1}.
       \label{c1half}
       \end{eqnarray}
cf.\  (\ref{eiam-series}), (\ref{defcoefC}).
The fermionic correlations are found in the form
       \begin{eqnarray}
       \langle
       \psi^\dagger_{R\sigma}(x,t)
       \psi_{R\sigma}(0)
       \rangle
       &\sim&
       \frac1{ (x-v_st)^{1/2} }
         \label{f-corr} \\ &\times &
       \sum_{n=-\infty}^\infty
       \frac{ c_n''\, e^{2\pi i\bar\rho x (n+1/4)}}
       {(x-v_F t)^{2n+1/2} 
       (x^2-v_F^2t^2)^{\gamma_n}}
      \nonumber
       \\
       \gamma_n &=&
       \frac{((2n+1/2){\cal K}-1)^2}{4{\cal K}}
       \end{eqnarray}
In the case of free fermions ($m=0$, $\beta^2=8\pi$) one 
has ${\cal K}=2$, and the anomalous dimension of the fermion 
vanishes, $\gamma_0 =0$. We observe that at special values of 
the Luttinger parameter, ${\cal K}/2 = 1/3, 1/5, 1/7, \ldots $, 
certain $\gamma_n$ vanishes as well, and the fermionic
correlations reveal purely chiral (subleading) component.  For
${\cal K} = 2/(2n+1)$, it is of the form

       \begin{equation}
       \frac{e^{ \pm \pi i\bar\rho x (n+1/2)}}
       {(x\mp v_F t)^{n+1/2} (x-v_st)^{1/2} }
       \end{equation}
with upper (lower) sign for even (odd) $n$.
The leading asymptote in (\ref{f-corr}) is given by a term with
$n=0$ and hence the scaling dimension of the fermion operator
has its usual value $dim[\psi_{R\sigma}] = \frac18 ( \sqrt{{\cal
K}/2} + \sqrt{2/{\cal K}})^2$.

\subsection{Low-density limit}

We have obtained the critical exponents for the correlation
functions which depend on the Luttinger parameter $\cal K$.
Some inspection of the formula (\ref{LuttParam}) reveals
that at small densities the Fermi velocity is exponentially
small, $v_F\simeq (m/2\bar\rho) e^{-m/2\bar\rho}$, and
the parameter ${\cal K} \sim \beta^2
(\bar\rho/m)^2 e^{m/2\bar\rho}$ eventually becomes large.
Haldane's analyzis of the Bethe-ansatz equations has shown
the crossover region at $4(\bar\rho /m) \ln \beta^{-1}\simeq1$
and the saturation of ${\cal K} \to 1$ at smaller densities.
This statement, albeit reasonable, is hard to check in our
formalism.

The saturation of $\cal K$ can be a result of the intervention
of some class of diagrams, increasingly important at small
densities. As we have shown, the full Lagrangian
(\ref{lagr-best}) is free from infra-red divergencies. On the
other hand, it requires the essential ultra-violet
regularization. The situation is reversed when one works with
the initial field variables (\ref{L-expa}).
Hence the question of saturation of $\cal K$ remains open
within our formalism.

\section{Discussion and conclusions}
\label{sec:conclu}

We have studied the sine-Gordon model
in the presence of a finite density of kinks.
This quantum problem is related to the strongly interacting
electronic system in one spatial dimension close to
commensurability, where the metal-insulator transition occurs.
The model was investigated previously by the Bethe ansatz method,
which gave a form of the spectrum and the Luttinger parameter,
defining the decay of correlations functions at large distances.

In dealing with complicated problems, it is always worth to have
alternative methods at hand, suitable for cross-checking with
existing results and capable for further theoretical predictions.
In this paper we used the semiclassical approach and demonstrated
that the spectrum and the Luttinger parameter correctly
reproduced the results of the Bethe ansatz studies. Among the new
features, available due to the quasiclassics, are the explicit
form of various correlation functions and the unambigous
construction of the charge-raising operator in the theory.

Discussing the correlation functions, we determined their leading
and subleading asymptotes, whose amplitudes and
critical exponents strongly depend on the kink density.
Addressing a more technical issue of the charge-raising operator,
or the dual field in the theory, we showed the way it is
constructed through the modified collective coordinate method.
This allowed us to re-establish the link with the fermionic
counterpart of our bosonic problem and to
calculate, in the leading order, the behavior of the
superconducting-type  correlations.

One the consequences of our approach is the non-zero value of
the quantum average, associated with the four-fermion umklapp
operator in the Hubbard model away from
half-filling. As a result, we found that close to
commensurability point, the spectral density of the right-
(left-) moving fermions is distributed over the neighboring
``$4k_F$'' harmonics of the fermionic density.

\acknowledgements

We thank D.I.Diakonov, A.Jerez, S.Lukyanov, N.Andrei,
D.L.Maslov for useful
discussions and communications. We are indebted to
V.V. Cheianov for numerous fruitful discussions.
D.A. acknowledges the partial financial support from
the grants RFBR 00-02-16873, FTNS 99-1134 and
the Russian State Program for Statistical Physics (Grant
VIII-2).

\appendix
\section*{collective coordinate method}
\label{sec:collect}

For brevity, we will write $\partial_t f=\dot f$, $\partial_x f=
 f'$
in this section.
Following \cite{ChristLee} we write our solution in the form
        \begin{equation}
        \phi(x,t) = \phi_0(x+X) +
        \sum_0^\infty c_n(t)\eta_n(x+X)
        \label{phiseries0}
        \end{equation}
The classical solution $\phi_0$ is proportional to $\beta^{-1}$,
 and the
fluctuations $\eta$ are of order of $\beta^0$. In terms of the
above mass of kinks condensate ${\cal M}_{tot}$,
the normalized zero-energy solution is
        \[
        \eta_0(x+X) = {\cal M}_{tot}^{-1/2} \phi_0'(x+X),
        \]
in accordance with (\ref{zeromode}). In order to pass to the
correct form of the quantum Hamiltonian, one has to regard the
variable $X$, which indicates the location of the soliton
lattice, as a dynamical variable, replacing the zero mode
coordinate $c_0$.  Instead of (\ref{phiseries0}) we write
        \begin{eqnarray}
        \phi(x,t) &=& \phi_0(x+X(t)) +
        \sum_1^\infty c_n(t)\eta_n(x+X(t))
        \label{phiseries1} ,
        \end{eqnarray}
and the time derivative
        \begin{eqnarray}
        \dot \phi(x,t) &=&
        (\phi'_0 +\sum_1^\infty c_n\eta_n')\dot X +
        \sum_1^\infty \dot c_n \eta_n
        \label{phidot1} ,
        \end{eqnarray}

Introducing the new coordinates $u_n, n=0,1,\ldots,\infty$
 according to
        $
        u_0(t)=X(t)$,  $u_n(t) = c_n(t) \quad {\rm for} \; n>0
        $
we represent the kinetic energy term in the Lagrangian in the
 form
        $
        \frac12\int_0^L dx\, \dot\phi ^2 =
        \frac12 \sum_{i,j=0}^\infty
        \dot u_i D_{ij}\dot u_j,
        $
with a symmetric matrix $D_{ij}$ with elements
        \begin{eqnarray}
        D_{00} &=& \int dx\,(\phi_0'
        +\sum_1^\infty u_n \eta_n' )^2 ,
        \nonumber \\
        D_{0n} &=& \int dx\, \eta_n
        (\sum_1^\infty u_m \eta_m' )
        , \quad n>0,
        \\
        D_{nm} &=& \delta_{nm}, \quad n,m>0.
        \nonumber
        \end{eqnarray}
The Hamiltonian is obtained by finding the canonical momenta
 $\pi_j$,
conjugate to coordinates $u_j$, according to
        $\pi_j = \partial {\cal L}/\partial \dot u_j
        = D_{ji}\dot u_i.$
The classical Hamiltonian then becomes
        \begin{eqnarray}
        {\cal H} &=&
        \frac12 \sum_{i,j=0}^\infty \pi_i (D^{-1})_{ij} \pi_j
        +V(\{ u_n\}),
        \label{ham0mode}
        \end{eqnarray}
where the potential energy term $V$ is given by
        $V(\{ u_n\}) =
        \int dx [\frac12 \phi'\,^2+ U(\phi)]=
        {\cal M}_{tot} +\frac12 \sum_{n=1}^\infty u_n^2\omega_n^2
 +
        O(\beta).$

The elements
of the inverse matrix $D^{-1}$ are
        \begin{eqnarray}
        (D^{-1})_{00} &=& 1/D ,
        \nonumber \\
        (D^{-1})_{0n} &=& -D_{0n}/D
        , \quad n>0,
         \\
        (D^{-1})_{nm} &=& \delta_{nm}+D_{0n}D_{0m}/D, \quad
 n,m>0.
        \nonumber
        \end{eqnarray}
and the determinant $D$ of the matrix $D_{ij}$ is given by
        $D = (\sqrt{{\cal M}_{tot}} + a_0)^2$ with
        $a_0 = \int dx\, \eta_0(x)
       (\sum_1^\infty u_m \eta_m')$.

Using above formulas, one obtains after some algebra
         \begin{eqnarray}
         \dot\phi(x,t) &=&  \frac{\eta_0}{\sqrt{D}} \pi_0 +
         \sum_{n=1}^\infty
         (\eta_n + \frac{D_{0n}}{\sqrt{D}}\eta_0 )\pi_n.
         \label{dotphi-mom}
         \end{eqnarray}

Requiring now
        $[u_n(t),\pi_m(t)] = i \delta_{nm}$
and using (\ref{dotphi-mom}),
one verifies the exactness of the equation
        $ [\phi(x,t),\dot \phi(y,t)]=i\delta(x-y)$,
which result improves the previous formula
(\ref{commut}).

Among the new features, brought about by the described method
is the explicit absence of the variable $X(t)$ in the
Hamiltonian (\ref{ham0mode}). As a result, its conjugate momentum
 $\pi_0$
is independent of time and equals to the conserved total field
 momentum
        $P \equiv \int_0^L dx\, \dot \phi \phi'$,
which equality is easily checked.

If we consider the soliton lattice in its rest frame, $P=0$, then
the Hamiltonian is simplified and becomes

        \begin{eqnarray}
        {\cal H}_{P=0} &=&
        \frac12 \sum_{n,m=1}^\infty
        (\delta_{nm}+
        \frac{D_{0n}D_{0m}}{D}) \pi_n \pi_m
        +V(\{ u_n\}),
        \label{ham0rest}
        \end{eqnarray}
At first glance, eqs. (\ref{dotphi-mom}), (\ref{ham0rest})
contain non-trivial admixtures $\propto D_{0n}/\sqrt{D}$, which
should interfere into the subsequent consideration.
This is true, when one deals with a vacuum, consisting of a
finite number of kinks $Q$. However, in the limit $Q\sim
L\to\infty$, one has $\eta_n \propto L^{-1/2}$ and therefore
$D_{0n}=O(1)$, whereas $D = O(Q \beta^{-2})$. As a result, the
terms $D_{0n}/\sqrt{D} \sim \beta Q^{-1/2}$ in
(\ref{dotphi-mom}), (\ref{ham0rest}) should be neglected in this
limit and we arrive at simpler equations ${\cal H}= {\cal E}_0
+ \frac12 \sum_{n=1}^\infty (\pi_n^2 + \omega_n^2 u_n^2) +
O(\beta)$ and
         $\dot \phi(x,t)_{P=0} =
         \sum_{n=1}^\infty \eta_n(x+X) \pi_n.
         $

\end{multicols}

\end{document}